\documentclass[]{zakoproc}
\usepackage{graphicx}
\begin{document}

\title{Magnetic fields and star formation as seen in edge-on galaxies}
\author{Marita Krause}
\institute{Max-Planck-Institut f\"ur Radioastronomie, Auf dem H\"ugel 69, 53121 Bonn, Germany}
\markboth{M. Krause}{Magnetic fields in edge-on galaxies \ldots}

\maketitle

\begin{abstract}
Radio continuum and polarization observations of several nearby galaxies allowed to determine their vertical scaleheights, magnetic field strengths and large-scale magnetic field patterns. They all show a similar large-scale magnetic field pattern, which is parallel to the galactic disk along the midplane and X-shaped further away from the disk plane, indepenent of their Hubble type or star formation in the disk or nuclear region. We conclude that -- though a high star formation rate (SFR) in the disk increases the total magnetic field strength in the disk and the halo -- the SFR does not significantly change the global field configuration nor influence the global scale heights of the radio emission. The observed similar scale heights indicate that star formation regulates the galactic wind velocities. The galactic wind itself may be essential for an effective dynamo action.
\end{abstract}

\section{Magnetic field strength and star formation}

Observations of three late-type galaxies with low surface-brightness and the radio-weak edge-on galaxy NGC~5907 (all with a low SFR) revealed an unusually high thermal fraction and weak total and ordered magnetic fields (Chy{\.z}y et al. 2007, Dumke et al. 2000). Even these objects follow the total radio-FIR correlation, extending it to the lowest values measured so far. Hence, these galaxies lack synchrotron emission compared to galaxies with higher SFR. Our findings, however, fit to the equipartition model for the radio-FIR correlation (Niklas \& Beck 1997), according to which the nonthermal emission increases $\propto SFR^{1.3 \pm 0.2}$ and the \emph{total}, mostly turbulent  magnetic field strength $ \rm B_t$ increases $\propto SFR^{0.34 \pm 0.14}$.\\
No similar simple relation is known for the \emph{ordered} magnetic field strength. We integrated the polarization properties in 41 nearby spiral galaxies and found that (independent of inclination effects) the degree of linear polarization is lower ($ < 4\%$) for more luminous galaxies, in particular those for $ L_{4.8} > 2 \times 10^{21}~\rm{W Hz^{-1}}$ (Stil et al. 2008). The radio-brightest galaxies are those with the highest SFR. Though dynamo action needs star formation and supernova remnants as the driving force for velocities in vertical direction, we conclude from our observations that stronger star formation seems to reduce the magnetic field regularity. On kpc-scales, Chy{\.z}y (2008) analyzed the correlation between magnetic field regularity and SFR locally within one galaxy, NGC~4254. While he found that the total and random field strength increase locally with SFR, the ordered field strength is locally uncorrelated with SFR.

\section{Vertical scale heights and CR-driven galactic wind}

We determined the vertical scale heights of the total power emission at $\lambda6$~cm of five edge-on galaxies (NGC~253, NGC~891, NGC~3628, NGC~4565, and NGC~5775) for which interferometer and single-dish data (VLA and the 100-m Effelsberg) were combined (Heesen et al. 2009, Dumke \& Krause 1998, Soida et al. 2011). In spite of the different intensities and extents of the radio emission, the vertical {\em scale heights} of the thin disk and the thick disk/halo are similar in this sample (see Table 1) with a mean value of $300 \pm 50$~pc for the thin disk and $1.8 \pm 0.2$~kpc for the thick disk. We stress that our sample includes the brightest halo observed so far, NGC~253, with strong star formation, as well as one of the weakest halos, NGC~4565, with weak star formation.

For NGC~253 Heesen et al. (2009) argued that the synchrotron lifetime (which is $\propto \rm B_t ^{-2}$) mainly determines the vertical scale height of the synchrotron emission and estimated the cosmic ray bulk velocity in NGC~253 to $300 \pm 30$~km/s. As this is similar to the escape velocity, it shows the presence of a galactic wind in this galaxy and implies that the galactic wind velocity is proportional to $\rm B_t ^2$. This may indicate that the total magnetic field strength and hence the star formation in the disk of a galaxy regulates the galactic wind velocity so that we see similar scaleheights in different galaxies.

\begin{table}[ht]
 \caption{Vertical scale heights at 4.8~GHz, star formation rates (SFR) as determined from IR-observations, star formation efficiencies (SFE), averaged total magnetic field strengths $\rm B_t$ in the disk, inclination i and Hubble-type for the edge-on spiral galaxies of our sample as far as the values could be determined up to now. References for the scales heights are given in the text. The Hubble type and i are taken from the literature.}
 \begin{center}
  \begin{tabular}[]{lccccccc}
          &\multicolumn{2}{c}{\textbf{Vertical scale heights}} & & & & &\\
   galaxy & thin disk & thick disk & \textbf{SFR}(IR) & \textbf{SFE} &  $\mathbf {B_t}$ & i & type\\
          & [pc] & [kpc] & [$ \rm{M}_\odot \rm{yr}^{-1} $] & [$ \rm{L}_\odot \rm{M}_\odot^{-1} $] & [$\mu$G] & [$\degr $]   & \\
   \hline
   NGC253 & $380 \pm 60$  & $1.7 \pm 0.1$ & 6.3 & 14 & 12 & 78 & Sc\\
   NGC891 & 270 & 1.8 & 3.3 & 5.0 & 6 & 88 & Sb\\
   NGC3628 & 300 & 1.8 & 1.1 & 4.9 & 6 & 89 & Sb pec\\
   NGC4565 & 280 & 1.7 & 1.3 & 3.2 & 7 & 86 & Sb\\
   NGC5775 & $240 \pm 30$ & $2.0 \pm 0.2$ & 7.3 & 6.1 & 8 & 86 & Sbc\\

   \hline
   mean    & $\mathbf {300 \pm 50}$ & $\mathbf {1.8 \pm 0.2}$\\
   \hline
   NGC4631 &    &      & 2.1 & 9.9 & 6 & 86 & SBd\\
   NGC5907 &    &      & 1.3 & 4.0 & 5 & 87 & Sc\\
   M104    &    &      & 1.2 & 4.2 & 4 & 84 & Sa \\
  \end{tabular}
 \end{center}
 \label{tableone}
\end{table}
\section{Magnetic field structure, dynamo action, and galactic wind}
In a larger sample of 8 edge-on galaxies (those of Table 1) we found in all of them (including NGC~4631 (Mora \& Krause, in prep.)) mainly a disk-parallel magnetic field along the galactic midplane together with an X-shaped field in the thick disk/halo as shown in Fig.1a for NGC~891. A similar field configuration was also found in NGC~4217 and NGC~4666 (Soida 2005), increasing the number to 10 galaxies. Our sample includes spiral galaxies of different Hubble types and SFR, ranging from about $ 0.6~\rm{M}_\odot \rm{yr}^{-1} \le \rm{SFR} \le 7.3~\rm{M}_\odot \rm {yr}^{-1}$ or correspondingly from star forming efficiencies from about $ 2~\rm{L}_\odot \rm{M}_\odot^{-1} \le \rm{SFE} \le 14~\rm{L}_\odot \rm{M}_\odot^{-1}$. The SFR-values were determined directly from the IR-emission given by Young et al. (1989) according to Kennicutt (1998), the SFE-values were determined from values for the molecular masses given in the literature.

\begin{figure}[htb]
\begin{minipage}[t]{6.0cm}
 \includegraphics[bb = 26 27 457 609, width=6.0cm,clip]{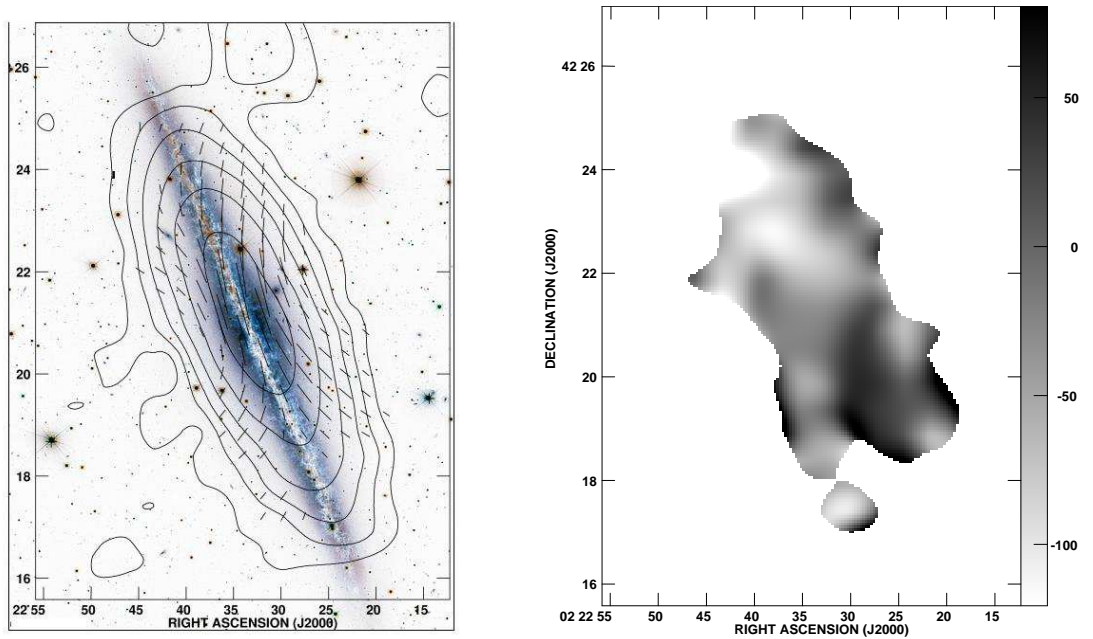}
\end{minipage}\hfill
\begin{minipage}[t]{7.2cm}
 \includegraphics[bb = 43 90 576 703, width=7.2cm, clip]{n891.RMintr.ps}
\end{minipage}
 \caption{(a) Total intensity (contours) of NGC~891 at 8.35~GHz with 84~$\arcsec$ HPBW resolution observed with the 100-m Effelsberg telescope (\copyright~MPIfR Bonn), overlayed on an optical image (\copyright~1999 CFHT/Coelum). The vectors give the intrinsic magnetic field orientation. (b) Observed RM $[\rm{rad/m}^2]$ of NGC~891 between 8.35~GHz and 4.75~GHz with $84\arcsec$~HPBW resolution.
}
 \label{description}
\end{figure}

The disk-parallel magnetic field is the expected edge-on projection of the spiral magnetic field within the disk as observed in face-on galaxies. Also the rotation measures RM in NGC~891 show an ordered, large-scale pattern along the major axis of the galaxy with negative values in the norh-east and positve values in the south-west. This indicates a coherent magnetic field pattern within the galactic disk. It is generally thought to be generated by a mean-field $\alpha \Omega$-dynamo for which the most easily excited field pattern is the axismmetric spiral (ASS) field (e.g. Beck et al. 1996). As there is no change in the sign of the RM visible {\em across} the major axis of NGC~891, the ASS field is of {\em even} parity in this galaxy. Taking into account that the north-eastern side of NGC~891 is the approaching side (e.g. from HI, Rupen 1991) (with RM beeing negative, hence the magnetic field is pointing away from us) and the south-western side is the receeding one (with RM beeing positive) we conclude that
the radial component of the ASS magnetic field points {\em outwards}. A similar result was obtained for NGC~5775 (Soida et al. 2011).

The poloidal part of the ASS dynamo field alone, however, cannot explain the observed X-shaped structures in edge-on galaxies as their field strengths seem to be comparable to that of the large-scale disk field. This is a factor of about 10 stronger than what the classical mean-field dynamo of the disk predicts for its poloidal component.

However, model calculations of the mean-field $\alpha\Omega$-dynamo for a disk surrounded by a spherical halo including a {\em galactic wind} (Brandenburg et al. 1993 and Moss et al. 2010) simulated similar magnetic field configurations to the observed ones. Meanwhile, MHD simulations of disk galaxies including a galactic wind implicitely may explain the X-shaped field (Gressel et al. 2008, Hanasz et al. 2009a,b). The first global, galactic scale MHD simulations of a CR-driven dynamo give promising results resembling the observations and show directly that small scale magnetic flux is transported from the disk into the halo (Hanasz et al. 2009c). A galactic wind can also solve the helicity problem of dynamo action (e.g. Sur et al. 2007). Hence, a galactic wind may be essential for an effective dynamo action and the observed similar X-shaped magnetic field structure in edge-on galaxies.

\end{document}